\documentclass[10pt,sigconf,letterpaper,nonacm,natbib=false]{acmart}

\usepackage{url}
\usepackage{numprint}
\usepackage[super]{nth}
\usepackage[inline]{enumitem}
\usepackage{todonotes}
\usepackage{siunitx}
\usepackage{float}
\usepackage{multirow}
\usepackage[htt]{hyphenat}

\graphicspath{{figures/}}

\AtBeginDocument{%
  }

\setcopyright{acmcopyright}
\copyrightyear{2018}
\acmYear{2018}
\acmDOI{XXXXXXX.XXXXXXX}

\RequirePackage[
  datamodel=acmdatamodel,
  style=acmnumeric,
  ]{biblatex}

\begin{document}

\title{Adaptive Address Family Selection for Latency-Sensitive Applications on Dual-stack Hosts}

\author{Maxime Piraux}
\email{maxime.piraux@uclouvain.be}
\orcid{0000-0002-5548-4423}
\affiliation{%
  \institution{UCLouvain}
  \country{Belgium}
}

\author{Olivier Bonaventure}
\email{olivier.bonaventure@uclouvain.be}
\orcid{0000-0002-6717-0296}
\affiliation{%
  \institution{UCLouvain}
  \country{Belgium}
}

\begin{abstract}
Latency is becoming a key factor of performance for Internet applications and has triggered a number of changes in its protocols. Our work revisits the impact on latency of address family selection in dual-stack hosts. Through RIPE Atlas measurements, we analyse the address families latency difference and establish two requirements based on our findings for a latency-focused selection mechanism. First, the address family should be chosen per destination. Second, the choice should be able to evolve over time dynamically.%

We propose and implement a solution formulated as an online learning problem balancing exploration and
exploitation. We validate our solution in simulations based on RIPE Atlas measurements, implement and evaluate
our prototype inside a DNS resolver in four access networks using Chrome and popular web services.
We demonstrate the ability of our solution to converge towards the lowest-latency address family and improve the latency of transport connections used by applications.
\end{abstract}

\maketitle

\section{Introduction}

The Internet originally adopted IPv4 as its network protocol. With the
important growth of the Internet over the years, the IPv4 address space became a
constraint that
the Internet Engineering Task Force (IETF) addressed by designing IPv6. It offers a much larger address space,
which also brings many opportunities~\cite{piraux2022multiple}. The Internet being a critical infrastructure and given the key role played by IP addresses, the
network community had to plan for the coexistence of IPv4 and IPv6 as a mean of
transitioning. This has led to a number of challenges in several areas of the
network stack. The DNS has been extended to resolve domain names into
IPv6 addresses with %
the
\texttt{AAAA}~\cite{rfc3596}
record type.
The IPv6 addressing scheme
comprises a dedicated prefix for
representing IPv4 addresses~\cite{rfc4291}, potentially simplifying the code of
applications supporting both.
The Linux socket API was also
modified %
with \textit{dual-stack sockets} enabling the use of IPv4 and IPv6, as well as
these IPv4-mapped IPv6 addresses. Windows and macOS APIs include methods to
connect a socket using domain name tackling this problem~\cite{ms-docs-connect,
apple-docs-connect}.

Given that the adoption of IPv6 on devices, operating systems and networks is
heterogeneous~\cite{nikkhah2015migrating, nikkhah2011assessing,
colitti2010evaluating}, very few service providers completely transitioned to
IPv6 but rather became dual-stack. As a result, when an application establishes a transport
connection%
, it needs to select one address family. This problem has seen a
number of solutions over the years~\cite{rfc6724, rfc6555, rfc8305}. All of them made the hypothesis that IPv6
should be favoured to foster its transition and include a fallback mechanism in
case of a broken IPv6 path. At the early stages of the IPv6 deployment, several
transition solutions such as 6to4~\cite{rfc3056} and Teredo~\cite{rfc4380}
introduced tunnels to carry IPv6 packets over IPv4. These techniques added
latency, could fail and become intermittently
unavailable~\cite{zander2012investigating}.
The first approach for applications was to start transport connections over
IPv6 and use IPv4 when
their establishment timer fires. Given that this time is usually set within
tens of
seconds, this approach greatly degrades the user experience when falling back to
IPv4. A refined algorithm named Happy Eyeballs (HE) races the
two connection
establishments while delaying the start of the IPv4 one, and selects the one
that completes first~\cite{rfc6555,rfc8305}. Its upcoming
proposed version~3~\cite{pauly-v6ops-happy-eyeballs-v3-01}
leverages \texttt{SVCB} DNS records~\cite{rfc9460,
zirngibl2023first} and honour the priority
they set among resolved addresses.
Today, as IPv6 is supported by \SI{48}{\%} of the 1000
top websites as reported by the ISOC Pulse
platform~\cite{isoc-pulse}, a significant part of applications
are facing this address family choice.

In the recent years, the greater importance in lowering the latency of Internet applications has started a series of changes in the Internet protocols. Two recent examples are the standardisation of the QUIC protocol~\cite{rfc9000}, with a faster handshake and better recovery mechanisms among its key features, and the L4S architecture~\cite{rfc9330}, which provides low latency under utilisation to applications using Active Queuing Management in the network. Both help reducing the round-trip time of packets between a host and a server.

In the light of these changes, we investigate and answer two questions in this paper: \begin{enumerate*}
    \item \textit{Does IPv6 always provide the same latency as IPv4 ?}
    \item \textit{How can we improve the address family selection for latency-sensitive applications ?}
\end{enumerate*}

Our paper addresses these questions using the following plan.
First, Section~\ref{sec:related} positions our work and research questions with respect to the state of the art.
Then, in Section~\ref{sec:e2e-latency-general}, we show, using
the RIPE Atlas
dataset~\cite{ripe-atlas}, that a diversity of latency
differences between
IPv4 and IPv6 exists. These differences can vary across several destinations
reached from a single source, with some also evolving over time.
Then, in Section~\ref{sec:design}, we formulate the lowest-latency address family selection as an online learning problem balancing exploration and exploitation. %
Section~\ref{sec:validation} validates our design with simulations using the RIPE Atlas dataset.
Section~\ref{sec:prototype} presents our prototype demonstrating how DNS
resolvers can aid hosts to select the lowest latency address family.
Section~\ref{sec:experiments} analyses results from real-world experiments
evaluating our prototype using Chrome and popular web services.
We will release all code from this work under an open-source
license~\cite{artefacts, updns} with the camera-ready version of this paper and
\appendixname~\ref{app:repro} details the steps to reproduce this work.

\section{Related works}
\label{sec:related}
Several works have studied the end-to-end performance of IPv6 that we categorise in three phases.
First, transition techniques such as 6to4~\cite{rfc3056} and
Teredo~\cite{rfc4380} that were prominent in the first phase of the worldwide
IPv6 deployment had a dual-sided impact, both fostering its deployment but also
often degrading user experience with added latency, sometimes poorer
availability and failures~\cite{guerin2010fostering}.
Second, with more native IPv6 networks, these effects gradually
decreased~\cite{czyz2014measuring}, and IPv6 and IPv4 had more often comparable
performance when the inter-domain paths were similar, while differing paths
significantly affected IPv6 performance~\cite{dhamdhere2012measuring}.
These routing differences have been established as both the biggest contributor to poor IPv6 performance at scale and a key point of improvement through better IPv6 peering~\cite{nikkhah2011assessing}.
Third, as the support of IPv6 in applications increased with adoption from
cloud companies and browsers, researchers analysed the impact of Happy Eyeballs
(HE) on performance. IPv6 TCP connections times have improved over time,
despite some still being higher than IPv4~\cite{bajpai2013measuring,
bajpai2015ipv4} which sometimes is due to lacking nearby IPv6 CDN
caches~\cite{bajpai2015ipv4}. HE was also found to
select IPv6 towards a large majority of popular web services despite IPv6 being
slower in most cases~\cite{bajpai2016measuring}. Halving the fallback timer of
HE to \SI{150}{ms} only enables marginal improvements in avoiding
higher-latency IPv6 connections~\cite{bajpai2016measuring,
bajpai2019longitudinal}.
In this work, we leverage the RIPE Atlas dataset~\cite{ripe-atlas} to assess the IPv4 and IPv6 performance differences today and validate our solution.

In the presence of dual-stack networks, researchers also proposed techniques
beyond HE. Given several available paths, multipath transport protocols such as
MPTCP~\cite{rfc8684, raiciu2012hard} and MPQUIC~\cite{ietf-quic-multipath-10,
de2017multipath} come as natural solutions. They can establish a connection
over one path and then a subflow on the other. Then, both be can used
simultaneously or selectively given application requirements, e.g. latency. An
extension to MPTCP proposed a way of racing the connection establishment over
several network paths and use all succeeding attempts within the MPTCP
session~\cite{amend2018robe}.
Researchers have also used MPTCP over IPv4 and IPv6 with a coupled congestion controller and demonstrated that bandwidth can be improved by aggregating the non-congruent parts of the two paths~\cite{livadariu2015leveraging}.
However, MPTCP is not widely used today~\cite{aschenbrenner2021single,
wang2023experience} while MPQUIC is still under
standardisation~\cite{ietf-quic-multipath-10}.

\section{IPv4 and IPv6 end-to-end Latency}%
\label{sec:e2e-latency-general}
We analyse the RIPE Atlas dataset~\cite{ripe-atlas} to investigate on the end-to-end latency of dual-stack networks.
RIPE Atlas consists of more than $10000$ probes which are connected, usually
using Ethernet, to networks around the world and measure connectivity with
several network tests. RIPE Atlas comprises also about $1000$ anchors which are
dedicated servers with more capacity used to co-operate with probes in their
measurements.
RIPE probes regularly perform HTTP tests towards anchors. %
This test opens a TCP connection, sends an HTTP/1.1 GET request
and receives a \SI{4}{KB} HTTP response in a single initial TCP
congestion
window. We use the request completion time of this test as a form of smoothed
latency measurement from a probe to a given anchor.

\begin{figure}
    \centering
    \includegraphics[width=\columnwidth]{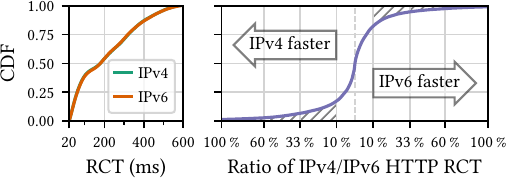}
    \vspace{-2em}
    \caption{Cumulative Distribution Functions (CDF) of HTTP
    request completion
    times (RCT) are similar
    across
    address families but RCT ratios have significant
    disparities when paired
    temporally.}
    \label{fig:ripe-http-ratio-dh}
\end{figure}

When a RIPE probe has access to a dual-stack network, the HTTP tests towards an
anchor are simultaneously performed over IPv4 and IPv6 within a short window of
time. When considering the HTTP
tests between the \nth{1} and \nth{8} of June 2023, we obtain
more than 156
millions of such paired tests, each happening within a
two-minute time window.
We filter out the completion times
outside the \nth{5} and \nth{95} percentiles.
The left part of \figurename~\ref{fig:ripe-http-ratio-dh}
reports the
Cumulative Distribution Functions (CDF) of HTTP request
completion times (RCT) for
both IPv4 and IPv6.
These are largely similar.
We also compute the completion time reach ratios of IPv4 over
IPv6 for each paired
test.

The right part of \figurename~\ref{fig:ripe-http-ratio-dh}
reports the
distribution of these ratios.
We can first observe that neither IPv4 nor IPv6 has a lower
completion time than
the other, which confirms our first finding. However, there is
an
important number of cases in which one of the two has a lower
completion time.
When looking at the \figurename, we can observe an equal
proportion of ratios have
a lower completion time by \SI{10}{\%} and more in favour of
one address family as
indicated by the hatched areas.
We can conclude that there is an important diversity in terms
of latency between
the IPv4 and IPv6 paths from a source to a destination.
When considering the use of Happy Eyeballs today, we observe that it can choose IPv6 when it has a higher latency than IPv4.
However, further analysis is required to answer two questions: \begin{enumerate*}
    \item \textit{How are these differences distributed over
    pairs of probe and anchor ?}
    \item \textit{How are these differences distributed over all anchors reached from a probe ?}
\end{enumerate*}

(1) First,
we group the HTTP tests data into about 400k pairs of probes and anchors for
which at least 300 HTTP tests are available.
We classify each pair into three
categories by performing a \textit{t-test} determining whether its IPv4 and
IPv6 RCT distributions have different means and compute a
\SI{98}{\%} confidence interval on the means difference. When the
\textit{p-value} is lower than $0.02$ and the confidence interval lower bound
is higher than the supposedly-higher distribution standard deviation, we define
the address family with a lower mean as better. When one of the two conditions
is not met, we define the pair has having no significantly better address
family. The second condition eliminates instances where the distributions
strongly overlap.
\begin{table}[]
    \centering
    \begin{tabular}{|c|c|c|}
        \hline
        IPv4 is better & IPv6 is better & None strongly better \\
        \hline
        113092 (28.4\%) & 129070 (32.4\%) & 156212 (39.2\%) \\
        \hline
    \end{tabular}
    \caption{Pairs of probe and anchor are rather homogeneously distributed
    between categories.}
    \label{tab:ripe-pairs-categories}
\end{table}
\tablename~\ref{tab:ripe-pairs-categories} reports the number of pairs in each
category. We can observe that while the dominant category contains pairs with
no significant difference, the others categories combined account for more than
60\% of pairs. While we considered pairs with tests spanning a week, we can
observe that our statistical tests establish a significant difference in a
majority of cases.

\begin{figure}
	\centering
	\includegraphics[trim={0pt 200pt 0pt
		0},clip,page=1]{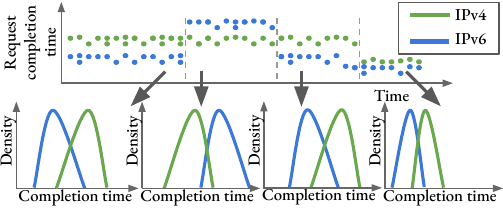}
	\vspace{-1em}
	\caption{An example of probe-to-anchor RCT series split
	into four segments
	using change-point detection
	to highlight dynamicity inside categories of
	\tablename~\ref{tab:ripe-pairs-categories} and produce
	\tablename~\ref{tab:ripe-pairs-segments}.}
	\label{fig:change-point-detection}
\end{figure}

Second, to better discern how stable these differences are, we deepen our
analysis by performing change-point
detection~\cite{wambui2015power} on the time series of each pair
of probe and anchor to split them into segments of at least 30 tests.
\figurename~\ref{fig:change-point-detection} illustrates this
process. We
then
perform our classification again on each segment to determine four additional
groups: \begin{enumerate*}
    \item[(A)] Pairs having a significantly better address family for which at least one segment is classified in the opposite address family.
    \item[(B)] Pairs having a significantly better address family for which a segment is classified as having no significantly better address family.
    \item[(C)] Pairs having no significantly better address family for which one segment is classified in one of them.
    \item[(D)] Pairs having all segments matching the global classification.
\end{enumerate*}

\begin{table*}
    \centering
    \small
    \begin{tabular}{|l|c|c|c|c|c|c|c|}
        \hline
        Group & \multicolumn{2}{c|}{(A)} & \multicolumn{2}{c|}{(B)} & \multicolumn{2}{c|}{(C)} & (D) \\
        \hline
        \tablename~\ref{tab:ripe-pairs-categories} category & IPv4 & IPv6 &
        IPv4 & IPv6 & None & None & \multirow{2}{*}{Consistent} \\
        $\exists$ segment with category & IPv6 & IPv4 & None & None & IPv4 &
        IPv6 & \\
        \hline
        Percentage & \multicolumn{2}{c|}{4569 (1.15 \%)} & \multicolumn{2}{c|}{32459 (8.15 \%)} & \multicolumn{2}{c|}{27312 (6.86 \%)} & 334034 (83.85 \%) \\
        \hline
    \end{tabular}
    \caption{While the classification of
    \tablename~\ref{tab:ripe-pairs-categories} remains
    consistent for 84\,\% of
    pairs,
    16\,\% diverge in category over time.}
    \label{tab:ripe-pairs-segments}
\end{table*}

\tablename~\ref{tab:ripe-pairs-segments} reports the number of pairs in each
group. We can observe that
more than 16 \% of the pairs of probe and anchor have time segments that
diverge in terms of best address family. Pairs can temporarily either see their
best address family change from one to the other (1.15 \% (A)), exhibit no
significant difference among the two
(8.15 \% (B)) or exhibit a significant difference for one of them (6.86 \%
(C)).
\textbf{We can
conclude that
differences in latency between IPv4 and IPv6 exist over a significant number of
pairs of probe and anchor. These differences can be dynamic and can fluctuate
over time}.

(2) Next, we investigate how the observed differences are spread among the
anchors reached by a probe. We group our classification of pairs per probe and
determine whether IPv4 or IPv6 performs the best in total across all
destinations, i.e. having the largest number of anchors reached with the lowest
latency. We then compute the ratio of anchors reached with this best address
family.
In the median case,
only \SI{39}{\%} of anchors can be reached with the lowest
RCT using this best address family alone. When adding anchors
having no strongly better address family, in the
median case, \SI{80}{\%} of anchors can be reached with the
lowest RCT.
\textbf{We can conclude that all dual-stack probes must use both address families to reach all anchors with the lowest completion time}.

\section{Adaptive Family Selection}%
\label{sec:design}

Based on our analysis presented in Section~\ref{sec:e2e-latency-general}, we observe two facts. First, IPv6 can provide a higher latency than IPv4 in a significant number of cases. Second, providing the lowest latency address family to applications cannot be achieved by entirely disabling one address family for a given client.
Moreover, we define two important properties for a latency-aware address family selection technique. First, the technique should be able to take different decisions for different destinations. Second, the technique must be able to adapt over time to changes in the address families latency.

\begin{figure*}
    \centering
    \includegraphics[trim={2pt 298pt 2pt
    0},clip,page=1]{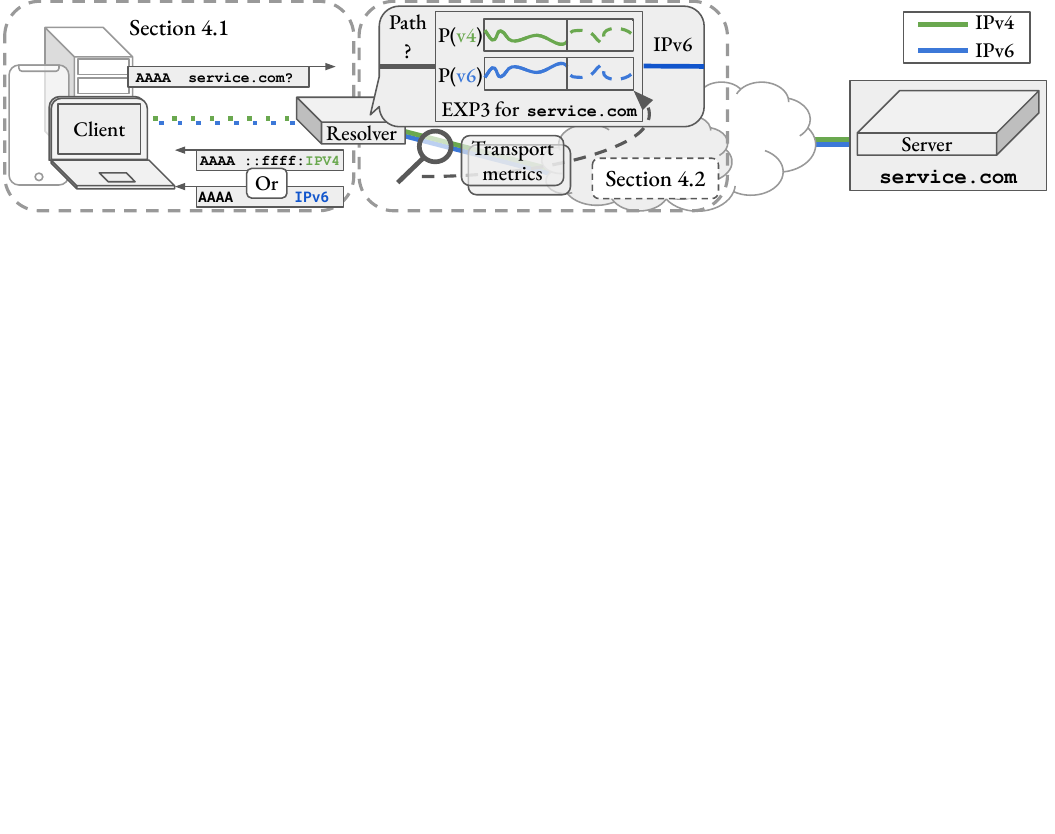}
    \vspace{-1em}
    \caption{A client in a dual-stack network using the DNS resolver
    selecting IPv6 as the best performing address family towards a given
    domain name based on an online learning algorithm fed by transport
    metrics.}
    \label{fig:dh-network}
\end{figure*}

This Section presents our design following these observations. \figurename~\ref{fig:dh-network} illustrates its application on a gateway within a home or small-enterprise network. The rest of the Section further elaborates on its components.

\subsection{Steering clients using the DNS resolver}

There is an opportunity for hosts to share their knowledge of RTTs
differences in their dual-stack network. %
We argue that the DNS resolver residing on the access gateway
could
influence the selected address family in its DNS responses.
In a dual-stack network such as illustrated on the left-hand side of \figurename~\ref{fig:dh-network}, the DNS resolver can choose the address family when answering to \texttt{AAAA} requests by using IPv6 and IPv4-mapped addresses. The latter allows encoding IPv4 addresses as IPv6 addresses using the dedicated \texttt{::ffff:/96} prefix~\cite{rfc4291}. These IPv6 addresses are not used on the wire and represent IPv4 addresses within the IPv6 address syntax.
Dual-stack applications are broadly compatible with these
addresses as the Linux kernel transparently supports them when
using the BSD socket API, as well as other operating systems.
An application using Happy Eyeballs will directly establish a
transport connection towards these addresses as they are
carried in \texttt{AAAA} records. %
Using HE v3~\cite{pauly-v6ops-happy-eyeballs-v3-01} and the
\texttt{SvcPriority} attribute of \texttt{SVCB} DNS records is
a more elegant solution but the former is not widely available
yet.

\subsection{Selecting the best address family}

When our DNS resolver receives a query in a dual-stack network, it has to decide which address family yields the lowest latency towards the requested domain name to choose which address it will answer from its cache. There are several ways of tackling this problem. A first approach is for the DNS resolver to asynchronously perform pings towards the requested domain name using both address families to inform its next choice towards this destination. It has several drawbacks, first it considers the ping time which can differ from the latency experienced by transport protocols due to network queuing and prioritisation%
. Second, it narrows the available information for the DNS resolver to choose the address family to a single metric, while metrics such as loss rates could be relevant to some latency-sensitive application. Third, it adds load on the network and destinations.

Another approach would leverage passive measurements, which alleviates these three drawbacks. Researchers have proposed efficient latency monitoring solutions that could be integrated in this work~\cite{soldani2023ebpf, sengupta2022continuous}. This approach can be taken by an on-path DNS resolver that would extract from TCP headers transport-level metrics such as connection establishment time and retransmissions. Off-path DNS resolvers could receive similar metrics using network telemetry. Based on this knowledge, the DNS resolver can discern which address family is better for each destination or prefix. For instance, a rolling history of past performance towards a given domain name could be used to select the most favourable address family. However, when steering the clients, the DNS resolver also has to explore the address families in order to rank them.

This tradeoff between exploration of the families and exploitation of the
best family is classical of multi-armed bandit problems, which are a class
of reinforcement learning problems. In our context, each address family is a
competing alternative action, as only one can be taken at a time, i.e. a
transport connection can only be established using one address family. Each
established connection can be then evaluated to estimate the gain associated
with using its address family towards a given destination by extracting
transport metrics (e.g, RTTs, loss rate, congestion window, $\ldots$).

\textbf{The EXP3 algorithm}.
The Exponential-weight algorithm for Exploration and Exploitation (EXP3) is
a
solution to this problem with several interesting
properties~\cite{auer2002nonstochastic, bubeck2012regret}.
First, it makes no statistical
assumptions on the process generating these gains. Second, it
is
resistant to adversarial models. Third, it was proven to converge towards
the
best action at a constant rate. Fourth, it is very computationally
lightweight
as its learning phase consists of a few float exponentiations and
multiplications. EXP3 sets probabilities on available actions and updates
them
based on the gain obtained for each action taken. Initial probabilities are
set equally. EXP3 defines a $\gamma$ parameter within the range $]0, 1]$
which
controls the balance between exploitation and exploration. When $\gamma =
1$,
EXP3 only explores actions but never considers what it learns, acting as a
random choice. Decreasing $\gamma$ decreases both the learning rate and the
exploration rate, i.e. the closer $\gamma$ is to 0, the slower EXP3 learns
but
the more it will restrict its exploration to rather exploit the best action.

The right-hand side of \figurename~\ref{fig:dh-network}
illustrates how an
instance of EXP3 estimates probabilities for each address family for a given
destination (P(v4) and P(v6)). These probabilities are updated based on the
gain observed from transport metrics for each taken action. They are used by
the DNS resolver to perform a probabilistic choice on the address family it
selects for a given destination.

\section{Validation}
\label{sec:validation}
\begin{figure*}
    \centering
    \includegraphics{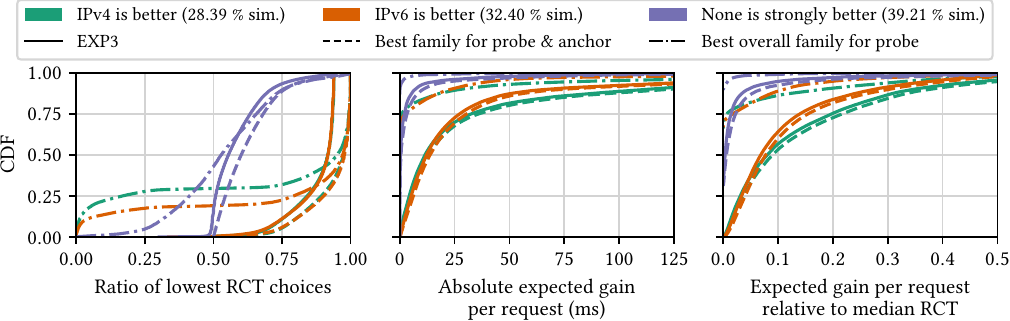}
    \vspace{-1em}
    \caption{EXP3 is largely able to choose the address family with the lowest completion time in simulations using the RIPE HTTP tests data from probes to anchors.}
    \label{fig:ripe-simulation}
\end{figure*}

In order to validate our solution in a controlled environment, we leverage the
RIPE data presented in Section~\ref{sec:e2e-latency-general} to test its
performance over a large amount of data. We extract more than 400k pairs of
probes and anchors with at least 300 IPv4 and IPv6 HTTP tests.
Given this large amount of probes
and anchors, we can use this data to evaluate our solution's ability to choose
the address family with the lowest request completion time (RCT) in a large
number of cases. To do so, we developed a Python simulator and an EXP3
implementation consisting of about 250 lines of code~\cite{artefacts}.

We define a simulation for each pair of probe and anchor consisting of one
round for each HTTP test from the probe to the anchor performed
simultaneously over IPv4 and IPv6. At the simulation start, we initialise a
EXP3 instance. A round consists of the following steps. First, the EXP3
instance draws the address family by forming a probabilistic choice. Second,
we record whether the RCT for
the chosen family is lower compared to the other family for this round.
Third, we compute the gain based on the chosen family to reward and train the
EXP3 instance using the following function: we give the entire reward for the
chosen address family when the obtained RCT is lower than the last RCT obtained
when EXP3 chose the other address family. Otherwise, we assign no reward to the
action.

We set the $\gamma$ EXP3 parameter to $0.1$, which makes EXP3 reach \SI{90}{\%}
probability for the best action within 60 rounds when it is strictly better
than the other action. When performing a simulation, we split the first 60
rounds for training and the rest for evaluation, during which we compute the
percentage of rounds where the family with the lowest RCT was chosen. We seed
the random number generator (RNG) once and run each simulation 100 times to
alleviate the RNG effect on the final score. We keep the median percentage of
the runs as the final score for each simulation.

\figurename~\ref{fig:ripe-simulation} reports the Cumulative Distribution
Functions (CDF) of three metrics computed from the results of our simulations.
We compare these three metrics in three categories
(\tablename~\ref{tab:ripe-pairs-categories}) across three methods.
These methods are:
\begin{enumerate*}
    \item[(plain lines)] our EXP3 approach,
    \item[(dashed lines)] a fixed choice picking the address family with the
    largest amount of lower RCT \textit{a posteriori} for this simulation,
    \item[(dashed and dotted lines)] a fixed choice picking the address family
    that performs the best for all simulations of this probe.
\end{enumerate*}
The second method acts as an upper bound on the impact of selecting the best address family in a simulation while the third highlights the impact of selecting the best address family as a whole for a given probe rather than per destination.

On the left-hand side of \figurename~\ref{fig:ripe-simulation}, we report the CDF of the median ratio of best choices, i.e. when the chosen address family leads to the lowest RCT, for each simulation.
We can observe that when a simulation exhibits a significant difference between the completion times of IPv4 and IPv6, EXP3 is largely able to converge towards the best address family and yields a high ratio of best choices. The distance between EXP3 and our higher bound selecting the best path a posteriori for the simulation is constant and can be explained by the exploration factor of EXP3. Reducing this distance could be obtained by lowering the $\gamma$ EXP3 parameter at the expense of a slower convergence.
When our statistical test establishes no significant difference between IPv4
and IPv6, we can observe that EXP3 is adversely affecting performance in a very
low number of simulations, i.e. simulations having a ratio less than $0.5$.
While EXP3 in most of these simulations gives a ratio lower than $0.55$, it is
also able to discern more subtle differences than established by
\textit{t-test} to select the best path. Indeed, the distance between EXP3 and
selecting the best path a posteriori for the simulation is similar to the other
categories.

When comparing EXP3 to our third method selecting the best path in average towards all destinations for a given probe, we observe an important regression for the latter with a different impact based on which family is best. When IPv4 exhibits a better performance, a higher percentage of simulations that choose the probe best path have strongly adverse performance. This can also be observed for IPv6 to a lower extent. This can be explained by several factors. First, IPv4 being better in a simulation is the rarest case, which means that for a probe, IPv4 being also the best path in average towards all destinations is rarest too. Second, when IPv4 is better, the distance from its mean completion time to IPv6 mean completion time is higher than in the opposite case. This could exacerbate the effect of the first factor.

To better understand the request completion time gain when selecting a given
path, we can observe the middle and right-hand side figures which depicts the
absolute expected gain per request and the gain relative to the
median RCT. When compared to selecting the best path a posteriori (dashed
lines), we can
observe that EXP3 yields a gain close to this upper bound. When comparing IPv4
and IPv6, we can observe that IPv4 has higher expected gain in a
small number of cases. When IPv4 has lower RCTs as established by our
statistical test, the
distance from the IPv4 to the IPv6 mean RCT is higher than in the
opposite case. We argue that this could be due to networks that are still
supporting IPv6 using transitions solutions introducing %
a greater overhead than the non-congruency of IPv4 and IPv6 paths alone.
When observing the right-hand side figure expressing this expected gain
as a ratio of the median RCT, we can observe that the gain yielded by EXP3 can
be higher than \SI{20}{\%} (resp. \SI{14}{\%}) of the median RCT for
\SI{25}{\%} of the cases when IPv4 (resp. IPv6) is best address family.
\textbf{We can conclude that EXP3 is able to converge towards the best address family yielding the lowest request completion time and resulting in a significant gain for a large number of pairs of probe and anchor.}

\section{Prototype}
\label{sec:prototype}
After validating the suitability of EXP3 using the RIPE Atlas dataset, we implemented our solution in a DNS proxy~\cite{updns} written in Rust. Our proxy receives DNS queries%
, queries an actual resolver and optionally alter its answers. We added 1000 lines of code to implement a DNS record cache, to improve the library used to manipulate DNS responses, and to implement EXP3 and our path selection technique.

When our prototype receives a DNS query, it uses EXP3 to probabilistically
choose which address family should be used. We took several implementation
decisions in this process. First, our prototype only acts when queries can be
resolved from its cache in order to avoid introducing additional delays during
the resolution process. When the cache is empty, our prototype simply acts as a
relay to an actual resolver and caches the received responses. Second, we group
in a single EXP3 instance destinations which share a common second-level domain
name. This balances the need for dedicating an EXP3 instance to each
destination with the benefit of grouping destinations experiencing similar
address family performance differences for accelerated learning. We make the
assumption that IP addresses resolved from domain names which share a common
second-level domain name are likely to experience similar performance
differences. Third, our prototype can also receive out-of-band feedback on the
performance of address families towards destinations via a custom UDP message
format. We use this feedback to compute EXP3 rewards using a similar function
as used in Section~\ref{sec:validation}: we give the full reward when the
address family used gave a better performance than the last use of the other
family, and none otherwise.

When receiving an \texttt{AAAA} query, our prototype leverages IPv4-mapped IPv6 addresses to select the address family to use towards the requested domain name. When an \texttt{A} query is received, our prototype respond with an empty answer to force the use of the selected address in the \texttt{AAAA} query. While the Happy Eyeballs algorithm as presented in Section~\ref{sec:background} recommends using the address received in the \texttt{AAAA} record, we found Chrome to wait for the \texttt{A} response to arrive.%

\section{Real-world experiments}
\label{sec:experiments}
We now evaluate our prototype using real networks and towards popular Internet services. While RIPE anchors are isolated servers to which a transport connection is established directly, popular services often have distributed infrastructures in several locations and place their servers behind load-balancers that can terminate the transport connections close to the users. We argue that measuring the end-to-end latency in this context for our work is relevant as it brings a different kind of destinations into perspective.

To find popular services, we use the Tranco list~\cite{LePochat2019} generated
on the \nth{5} of April 2023~\cite{tranco-url}
and choose the top 40 domain names for which an IPv4 and IPv6 address can be
resolved from our campus network. The resulting list has a diversity of
companies and organisations providing popular services to a significant part of
the Internet users.

We use the browserless Docker image~\cite{browserless} to evaluate the impact
of our prototype on Chrome performance in a reproducible environment. We
configure Chrome to use our prototype as a DNS resolver, which leverages the
DNS resolver within the network where experiments are performed. During an
experiment, our prototype is started without any prior knowledge and Chrome is
instructed to load websites one at a time from a list based on the top 40
domain names. To measure our prototype ability to learn and exploit the best
address family towards destinations, we repeat in the list each domain name 30
times. The list is then shuffled to avoid bias. When running the experiment, we
load each website from the list in three modes: IPv4 only, IPv6 only, and our
prototype. This gives us a reference point for each address family to compare
the quality of the choices of our prototype. These modes are also run in random
order and using separate instances such that no previous run from other modes
influences our prototype nor the browser. After a website is fully loaded using
our prototype as a DNS resolver, we extract the handshake time of all transport
connections established for the website. Then, the rewards for EXP3 are
computed and attributed to the corresponding domain names using the same
formula as described in Section~\ref{sec:validation}.%

We perform this experiment from several vantage points in Belgium: \begin{enumerate*}
    \item our university campus network,
    \item a DSL line,
    \item a cable line,
    \item a 4G Fixed Wireless Access network
\end{enumerate*}. All four networks are operated by different ISPs. When
measuring from a vantage point, the Linux test device is connected using
Ethernet. We collect all transport connections handshake times during these
experiments. Then, we process them such that the first 60 tests towards a
destination are not considered as they correspond to the training phase of our
prototype. Then we keep the destinations for which at least 100 tests exist,
such that a significant amount of connections was steered using our trained
prototype. Finally, to avoid over-representing websites that use more transport
connections than others, and destinations that are more frequent among these
websites, such as common CDN resources, we compute the mean handshake time for
each destination. We repeat this process for the two other modes.

\begin{figure}
    \centering
    \includegraphics{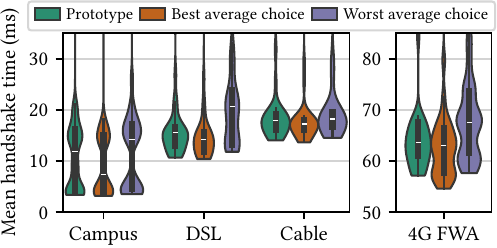}
    \vspace{-2em}
    \caption{Our prototype is largely able to choose the best address family
    in experiments using four access networks %
    when loading popular websites with Chrome.}
    \label{fig:browserless}
\end{figure}

\begin{table}
	\centering
	\Small
	\begin{tabular}{|l|c|c|c|c|}
		\hline
		& Campus & DSL & Cable & 4G FWA \\
		\hline
		Prototype & \SI{11.8}{ms} & \SI{15.5}{ms} & \SI{17.9}{ms} &
		\SI{63.6}{ms} \\
		Best & \SI{7.3}{ms} & \SI{14.2}{ms} & \SI{17.2}{ms} & \SI{63.1}{ms} \\
		Worst & \SI{14.3}{ms} & \SI{20.7}{ms} & \SI{18.2}{ms} & \SI{67.5}{ms}
		\\
		\hline
	\end{tabular}
	\caption{Medians reported in \figurename~\ref{fig:browserless}.}
	\label{tab:real-world-median}
\end{table}

\figurename~\ref{fig:browserless} reports the distributions of the mean
handshake time per destination when loading these popular websites using our
prototype on these four access networks. We compare them to the best and worst
address families for each destination. We determine a posteriori which address
family is better or worse using the handshake times experienced in the IPv4 and
IPv6 only modes. When looking first at each distribution and
\tablename~\ref{tab:real-world-median}, we can observe that our prototype
improves the mean handshake time by selecting the best address family for each
destination. In addition, the tail of the distributions is also reduced. There
is no network in which our prototype degrades the handshake times. Its mean
handshake time distribution is close to the ideal solution, and its small
deviation can be explained by the EXP3 exploration. When looking at each
network,
the DSL line experiences the largest differences in terms of address family
performance.

\section{Discussion and further work}
This work opens several directions for further work.
First, we argue that the DNS protocol should be extended to allow indicating the address family %
that a host should use with a dedicated additional record. The Happy Eyeballs algorithm could be updated to take those hints into account when sorting the received addresses.
Second, the problem of address family selection in dual-stack networks being a
subset of the path selection problem in multihomed networks, our approach could
be extended to IPv6-multihomed networks. An enterprise subscribed to several
network providers could use its DNS resolver to hint the source prefix to use
towards a given destination.
Finally, our approach could leverage other transport metrics than the connection handshake time. For instance, some latency-sensitive applications such as video conferencing also require a low loss rate and jitter while others might require a more stable bandwidth.

\clearpage

\begin{acks}
We thank Louis Navarre and François Michel for their comments and feedback that helped improve this work.
\end{acks}

\printbibliography

\appendix
\section{Happy Eyeballs}
\label{sec:background}

Starting during the early deployment phase of IPv6 and its start of coexistence with IPv4, IPv6-enabled applications faced the choice of the address family when establishing a transport connection to a domain name that resolves in both families. While IPv6 could be preferred to favour the IPv6 deployment, network paths where it could be failing or be largely inferior existed.
The approach taken by the IETF to solve this problem is to provide to
applications an address selection mechanism favouring IPv6 that could quickly
fall back in case of unexpected connection delay. The algorithm is called Happy
Eyeballs and was first standardized in RFC~6555~\cite{rfc6555}, and then
revised in RFC~8305~\cite{rfc8305}.

\begin{figure}
    \centering
    \def\svgwidth{\columnwidth}
    \input{figures/happy_eyeballs_v2_t.tex}
    
    \caption{Happy Eyeballs Version 2 in RFC~8305.}

    \label{fig:he-v2}
\end{figure}

\figurename~\ref{fig:he-v2} illustrates the state machine of the Happy Eyeballs Version 2 algorithm as described in RFC~8305. The algorithm introduces several choices. IPv6 is favoured when choosing the first address to establish a transport connection. When the DNS resolution of IPv6 is delayed compared to IPv4 (state 3), the algorithm recommends waiting \SI{50}{ms} before trying the IPv4 address first. When an empty AAAA DNS record is received, IPv4 is tried as soon as possible. When an address is tried, the algorithm specifies three time bounds before trying another address from another family. The next attempt delay must not be lower than \SI{10}{ms}, with a recommended value of \SI{100}{ms} and a maximal recommended value of \SI{2}{s}.

\section{Reproducibility}
\label{app:repro}
This \appendixname\ details the required steps and material to
reproduce the results presented in this work. All the code
developed and data referenced in this section will be made
publicly available under an open-source
licence~\cite{artefacts, updns}. An archive containing those
two repositories was submitted along with this paper.
The RIPE data that we downloaded and processed is subject to
their legal terms~\cite{ripe-terms}.

\subsection{IPv4 and IPv6 end-to-end latency}
\label{app:e2e}

The RIPE data presented was first extracted using the BigQuery interface~\cite{ripe-bq}. The query used is in the file \texttt{/e2e\_latency/query.sql}. The resulting data should be saved in a CSV file. For this purpose, the helper Python script \texttt{/e2e\_latency/download\_bq\_results.py} connects to BigQuery and downloads a \texttt{data.csv} file. It must be adapted with credentials and BigQuery Job ID. The expected file size is around \SI{13.4}{GB}. The \texttt{/e2e\_latency/figure\_1.py} script produces \figurename~\ref{fig:ripe-http-ratio-dh}. We ran it on machine with \SI{128}{GB} of RAM as it loads the entire CSV file.

To produce Table~\ref{tab:ripe-pairs-categories},
\ref{tab:ripe-pairs-segments} and the numbers cited in (2) of
Section \ref{sec:e2e-latency-general}, the required data is
produced by the Python
simulator described in Section~\ref{sec:validation}. The rest
of the output of the simulator is discussed in the other parts
of the \appendixname. The simulator use a condensed view of the
data, so the \texttt{data.csv} must be processed into the
\texttt{paths.cbor} file using the
\texttt{/simulations/prepare\_data.py} Python script. This
script was also run on machine with \SI{128}{GB} of RAM. Then,
running the \texttt{/simulations/ripe.py} Python script
produces the required output. The rest of the Python scripts in
the \texttt{/e2e\_latency} folder (i.e  \texttt{table\_1.py},
\texttt{table\_2.py})
can be run to
obtain the corresponding Table.

\subsection{Validation}

\figurename~\ref{fig:ripe-simulation} requires running first
the simulator as described in \appendixname~\ref{app:e2e}. Then
\texttt{/validation/figure\_4.py} can be run to produce the
\figurename.

\subsection{Real-world experiments}

The \texttt{/experiments} folder contains the scripts necessary to run experiments and parse the results into the JSON format used to produce \figurename~\ref{fig:browserless}.
The real-word experiments require a Linux machine with Docker installed. We used the \texttt{browserless/chrome} Docker image hosted on \texttt{docker.io} with ID \texttt{e6460f3f8f60}.  Our prototype is available in a separate repository~\cite{updns}. Its Docker image should be built before running experiments. Then, the \texttt{/experiments/run\_browserless.py} Python script starts all the experiments described Section~\ref{sec:experiments} over a given network. This script needs to be adapted to specify the interface to use through indicating its IPv4 address in the \texttt{IP} global variable.
The results are stored invidually in a directory. This directory can be parsed by the \texttt{/experiments/parse\_netlogs.py} Python script into a JSON file format summarising all results. The \texttt{/experiments/figure\_5.py} Python script illustrates how \figurename~\ref{fig:browserless} was produced and should be modified when the number of networks tested changes.

\end{document}